\documentstyle[preprint,aps]{revtex}

\input psfig.sty

\begin{document}
\draft
\title{Collective neutrino-pair emission due to Cooper pairing of protons in
superconducting neutron stars}
\author{L. B. Leinson\footnote{Permanent address:
Institute of Terrestrial Magnetism, Ionosphere and Radio Wave Propagation
RAS, 142092 Troitsk, Moscow Region, Russia.}}
\address{Departamento de F\'{i}sica Te\'{o}rica, Universidad de Valencia\\
46100 Burjassot (Valencia), Spain.}
\maketitle
\begin{abstract}
The neutrino emission due to formation and breaking of Cooper pairs 
of protons in superconducting cores of neutron stars is considered 
with taking into account the electromagnetic coupling of protons to 
ambient electrons. It is shown that collective response of electrons 
to the proton quantum transition contributes coherently to the complete
interaction with a neutrino field and enhances the  neutrino-pair 
production. Our calculation shows that the contribution of the vector 
weak current to the $\nu \bar{\nu}$ emissivity of protons is much larger 
than that calculated by different authors without taking into account 
the plasma effects. Partial contribution of the pairing protons to the 
total neutrino radiation from the neutron star core is very sensitive 
to the critical temperatures for the proton and neutron pairing. We show 
domains of these parameters where the neutrino radiation, caused by a 
singlet-state pairing of protons is dominating.

\end{abstract}

\widetext
\section{Introduction}

\label{sec:intro}

When the temperature inside a neutron star core is lower than the critical
temperature $T_{c}$ for nucleon pairing, the nucleon matter exhibits a
condensate of Cooper pairs, which has thermal excitations in the form of not
paired quasi-particles. Cooper-pair formation and pair-breaking coexist in
statistical equilibrium and result in additional neutrino-pair emission from
the neutron star. Under certain conditions, neutrino emission due to Cooper
pairing of nucleons may be sufficiently high, compared with or even larger
than the emissivity of Modified Urca process in non-superfluid matter. This
mechanism of energy loss was proposed by Flowers et al. many years ago \cite
{FRS76} but have been included into cooling simulations only recently \cite
{Vosk97}, \cite{Page98}, \cite{Yak98}. It was shown, that neutrino emission
due to formation and breaking of Cooper pairs can accelerate the neutron
star cooling. 

Flowers et al. \cite{FRS76} have studied the neutrino-pair emissivity for
the case of a singlet-state pairing of neutrons. One more independent
calculation of this process was suggested by Voskresensky and Senatorov \cite{52}. 
They have obtained a similar equation for the neutrino-pair emissivity,
but contrary to previous authors, the expression obtained by Senatorov and
Voskresensky contains the axial-vector contribution. A new 
calculation, performed by Yakovlev et al. \cite{YKL98}, has confirmed that 
the axial-vector contribution arises actually due to 
relativistic effects and is negligible in the case of 
singlet-state pairing of neutrons. 

Yakovlev et al. \cite{YKL98} have considered
also the more probable case of a triplet-state neutron pairing. 

The case of proton pairing was considered in \cite{KHY99}. 
Specifics of the proton pairing, which we are going to discuss in this paper,
occurs due to a smallness of the weak vector coupling of a proton. Cooper
pairing of protons take place likely in $^{1}S_{0}$-state \cite{TT93}. When
protons are treated non-relativistically, the total spin of the Cooper pair
in the singlet-state is zero. By this reason, the axial-vector
contribution of the proton weak current to the neutrino emissivity occurs
as a relativistic correction, which, however, is found to be
larger than the vector contribution due to the numerical smallness of the 
above weak vector constant of protons\cite{KHY99}. 

In our paper we argue that, in spite of the smallness of the proton vector
weak coupling, the latter plays an important role in the neutrino-pair
radiation from pairing protons. The above inference about negligible
contribution of the vector weak coupling was made on the basis of
calculations which ignored electromagnetic correlations among the charged
particles in a QED plasma. Actually, protons in the plasma are coupled to
ambient electrons via the electromagnetic field. The electron vector weak
coupling with a neutrino field is much stronger than that for protons,
therefore the virtual electron cloud, associated to the medium polarization,
strongly modifies the effective vector weak current of protons \cite{L2000}.
The collective effects caused by 
polarization of the electron plasma manifest themselves in the Debye volume
around the proton. As the wavelength of radiated neutrino and antineutrino
is much larger than the electron Debye screening distance $D_{e}$
(Typically, the ratio is of the order of 10 or larger.), the induced virtual 
electron
excitations generate neutrinos coherently with the protons undergoing the
quantum transition and leads to stronger neutrino emission than that 
due to the direct proton interaction with a neutrino field. 

In the present article we suggest the calculation of $\nu \bar{\nu}$
radiation caused by Cooper pairing of protons in the $^{1}S_{0}$-state,
taking into account the above collective effects. 
We consider the medium as consisting
of neutrons, protons, and electrons. Generalization including
muons is also considered. 

To incorporate the electromagnetic correlations, one has to take into account
exchange of photons between charged particles in the medium. For this we 
first consider some properties
of the electromagnetic field in the plasma consisting of superconducting
protons and normal degenerate electrons. In Sec.\ \ref{sec:General} we
discuss the corresponding electromagnetic field equations, polarization
functions of the medium, and derive the in-medium photon propagator. Further
we demonstrate the Lagrangian to be used for weak interaction of protons and
electrons with a neutrino field. Here we discuss also 
the problem of nuclear renormalization of the weak proton vertex. 
In Sec. \ref{sec:emis} we derive the general formula for
neutrino-pair emissivity by the use of the Optical theorem. 
In Sec.\ \ref{sec:loop} we apply the closed
diagrams to calculate the emissivity of the process in the loop
approximation reproducing the result obtained by different authors,
without taking into account the collective effects. In Sec.\ \ref{sec:RPA}
we calculate the neutrino-pair emissivity by the use of the Random phase
approximation, which incorporates the electromagnetic correlations in the
medium. We derive the expression for the RPA neutrino-pair emissivity.
Some practical formulae and numerical estimates are given in Sec.\ 
\ref{sec:numer}. In Sec.\ \ref{sec:effic} we show numerical results for some
typical parameters of nuclear matter, and discuss the efficiency of
neutrinos due to proton pairing. Summary and conclusion are in Sec.\ \ref
{sec:concl}.

In what follows we use the system of units in which $\hbar =c=1$ and the
Boltzmann constant $k_{B}=1$.

\section{General formalism}
\label{sec:General}
To incorporate the collective plasma effects, one should take into account
exchange of photons between charged particles in the plasma. For this we
consider first the electromagnetic properties of the medium consisting of
superconducting protons and normal degenerate electrons.

\subsection{Electromagnetic field equations}

The superconducting electric current of protons and the normal electric
current of electrons coexist, participating in electromagnetic oscillations
of the medium. At zero temperature, the total Lagrangian density of both the
complex field $\Psi $ of proton Cooper pairs and electrons interacting with
the electromagnetic field, has the form 
\begin{equation}
L=\left| \left( \partial _{\mu }+2ieA_{\mu }\right) \Psi \right| ^{2}+M_{%
\tilde{C}p}^{2}\left| \Psi \right| ^{2}-\kappa \left| \Psi \right| ^{4}-%
\frac{1}{16\pi }F_{\mu \nu }^{2}-j_{\mu }^{e}A^{\mu }+L_{e}^{0},
\label{Ldens}
\end{equation}
where $\kappa \ $is a constant for proton-proton interaction resulting in
the proton pairing, and $M_{\tilde{C}p}\simeq 2M^{\ast }$ is the mass of the
Cooper pair consisting of two protons of effective mass $M^{\ast }$; $F_{\mu
\nu }=\partial _{\mu }A_{\nu }-\partial _{\nu }A_{\mu }$ is the tensor of
electromagnetic field; $j_{e}^{\mu }$ is the electron current, and $%
L_{e}^{0} $ is the Lagrangian density of free electrons. As a result of
spontaneous breaking of the symmetry of the ground state of the system, the
vacuum expectation value of the Cooper pair field is nonzero $\left\langle
\left| \Psi \right| ^{2}\right\rangle $ =$\Psi _{0}^{2}$. Since the ground
state of the system corresponds to zero spin and zero total momentum of the
Cooper pair the vacuum expectation value is connected with the number
density of Cooper pairs $N_{s}$ by the relation $N_{s}=2E\Psi _{0}^{2}$,
where $E\approx 2M^{\ast }$ is the energy of the Cooper pair corresponding
to zero total momentum, and $N_{s}$ is one-half the number density of paired
protons $N_{p}$. Thus, 
\begin{equation}
\Psi _{0}^{2}=\frac{N_{p}}{8M^{\ast }}.  \label{1}
\end{equation}
With allowance for the interacting with the electromagnetic field $A_{\mu }$%
, we represent the field of the Cooper condensate in the form 
\begin{equation}
\Psi =\Psi _{0}\left( 1+\rho \right) \exp \left( -2ie\phi \right) ,
\label{2}
\end{equation}
where $\rho $ and $\phi $ are arbitrary real functions of the coordinates
and time, and substitute in the Lagrangian density (\ref{Ldens}) for the
fields $\rho $ and $\phi $. Taking into account the following identity 
\begin{equation}
\left| \left( \partial _{\mu }+2ieA_{\mu }\right) \Psi \right| ^{2}=\Psi
_{0}^{2}\left( \partial _{\mu }\rho \right) ^{2}+4e^{2}\Psi _{0}^{2}\left(
A_{\mu }-\partial _{\mu }\phi \right) ^{2}\left( 1+\rho \right) ^{2},
\label{3}
\end{equation}
we can make the gauge transformation $A_{\mu }^{\prime }=A_{\mu }-\partial
_{\mu }\phi $. Since the quantity $F_{\mu \nu }=F_{\mu \nu }^{\prime
}=\partial _{\mu }A_{\nu }^{\prime }-\partial _{\nu }A_{\mu }^{\prime }$ as
well as the electron current $j_{\mu }^{e}=j_{\mu }^{\prime e}$ are
gauge-invariant we obtain

\begin{equation}
L=\psi _{0}^{2}\left( \partial _{\mu }\rho \right) ^{2}+4e^{2}\Psi
_{0}^{2}A_{\mu }^{\prime 2}\left( 1+\rho \right) ^{2}+M_{\tilde{C}%
p}^{2}\left| \psi \right| ^{2}-\lambda \left| \psi \right| ^{4}-\frac{1}{%
16\pi }F_{\mu \nu }^{\prime 2}-j_{\mu }^{e}A^{\prime \mu }+L_{e}^{0}.
\label{4}
\end{equation}
As a result, the Goldstone field $\phi $ is absorbed by the gauge
transformation. Considering this and taking variations of the Lagrangian
with respect to the field $A^{\prime \mu }$, in the linear approximation we
obtain the equation 
\begin{equation}
\partial _{\nu }\partial ^{\nu }A_{\mu }^{\prime }+32\pi e^{2}\Psi
_{0}^{2}A_{\mu }^{\prime }=4\pi j_{\mu }^{e},\text{ \ \ \ }\partial ^{\mu
}A_{\mu }^{\prime }=0.  \label{A}
\end{equation}

In the absence of the electron current $\left( j_{\mu }^{e}=0\right) $, this
equation would describe the eigen photon modes of mass 
\begin{equation}
m_{\gamma }=\sqrt{32\pi e^{2}\Psi _{0}^{2}}=\sqrt{4\pi N_{p}e^{2}/M^{\ast }}.
\label{mg}
\end{equation}
This is the well-known Higgs effect. The Goldstone field $\phi $ has been
absorbed by the gauge transformation. As a consequence of this, the photon
has acquired a mass and an additional longitudinal polarization.

However, the polarization of the electron plasma modifies the massive photon
spectrum. In fact, in the four-momentum representation, the electron
current, induced by the electromagnetic field, is $4\pi j_{\mu }^{e}=-\pi
_{\mu \nu }^{\left( e\right) }A^{\prime \nu }$. Here $\pi _{\mu \nu
}^{\left( e\right) }$ is the electromagnetic polarization tensor of the
electron plasma. At finite temperature, one has to take also into account
the electric current of the normal proton component consisting of non-paired
proton quasi-particles. This can be described by the proton electromagnetic
polarization tensor $\pi _{\mu \nu }^{\left( p\right) }$. One has $4\pi
j_{\mu }^{p}=-\pi _{\mu \nu }^{\left( p\right) }A^{\prime \nu }$. Thus, the
equation for the field (\ref{A}) takes the form (In what following we omit
prime in the notation of the field): 
\begin{equation}
K^{2}A_{\mu }-m_{\gamma }^{2}A_{\mu }-\pi _{\mu \nu }\left( K\right) A^{\nu
}=0,\text{ \ \ \ }K^{\mu }A_{\mu }=0,  \label{EqA}
\end{equation}
where the electromagnetic polarization tensor is: 
\begin{equation}
\pi _{\mu \nu }=\pi _{\mu \nu }^{\left( e\right) }+\pi _{\mu \nu }^{\left(
p\right) }  \label{5}
\end{equation}

\subsection{Orthogonal basis}

To specify the components of the polarization tensors, we select a basis
constructed from the following orthogonal four-vectors $\ $ 
\begin{equation}
h^{\mu }\equiv \frac{\left( \omega ,{\bf k}\right) }{\sqrt{K^{2}}},\text{ \
\ \ \ }l^{\mu }\equiv \frac{\left( k,\omega {\bf n}\right) }{\sqrt{K^{2}}},
\label{kh}
\end{equation}
where the space-like unit vector ${\bf n=k}/k$ is directed along the
electromagnetic wave-vector ${\bf k}$. Thus, the longitudinal basis tensor
can be chosen as $L^{\rho \mu }\equiv -l^{\rho }l^{\mu }$ , with
normalization $L_{\rho }^{\rho }=1$. The transverse (with respect to ${\bf k)%
}$ components of $\Pi ^{\,\rho \mu }$ have a tensor structure proportional
to the tensor $T^{\rho \mu }\equiv \left( g^{\rho \mu }-h^{\rho }h^{\mu
}+l^{\rho }l^{\mu }\right) $, where $g^{\rho \mu }={\sf 
\mathop{\rm diag}%
}(1,-1,-1,-1)$ is the signature tensor. This choice of $T^{\rho \mu }$
allows us to describe the two remaining directions orthogonal to $h$ and $l$%
. Therefore, the transverse basis tensor has normalization $T_{\rho }^{\rho
}=2$. One can also check the following orthogonality relations: $l_{\rho
}T^{\rho \mu }=0,$ as well as $k_{i}T^{i\mu }=0,$ and $\ K_{\rho }L^{\rho
\mu }=K_{\rho }T^{\rho \mu }=0$.

In this basis 
\begin{equation}
\pi ^{\,\rho \mu }\left( K\right) =\pi _{l}\left( K\right) L^{\rho \mu }+\pi
_{t}\left( K\right) T^{\rho \mu },  \label{pi}
\end{equation}
where the longitudinal polarization function is defined as $\pi _{l}\left(
K\right) =\left( 1-\omega ^{2}/k^{2}\right) \pi ^{\,00}$ and the transverse
polarization function is found to be $\pi _{t}\left( K\right) =\left(
g_{\rho \mu }\pi ^{\rho \mu }-\pi _{l}\right) /2$.

\subsection{Real part of polarizations}

To the lowest order in the fine structure constant, the polarization
tensor of the electron gas is defined as follows 
\begin{equation}
\pi ^{\left( e\right) \,\mu \rho }\left( K\right) =4\pi ie^{2}%
\mathop{\rm Tr}%
\int \frac{d^{4}p}{(2\pi )^{4}}\,\gamma ^{\mu }\,{\hat{G}}(p)\,\gamma ^{\rho
}\,{\hat{G}}(p+K),  \label{PT}
\end{equation}
Here ${\hat{G}}(p)$ is the in-medium electron Green's function. This is
so-called one-loop approximation corresponding to a collisionless electron
plasma.

In the case of a strongly degenerate ultrarelativistic electron plasma $%
\left( v_{F}\simeq 1\right) $, one has \cite{BS93}: 
\begin{equation}
\pi _{l}^{\left( e\right) }=\frac{1}{D_{e}^{2}}\left( 1-\frac{\omega ^{2}}{%
k^{2}}\right) \left( 1-\frac{\omega }{2k}\ln \frac{\omega +k}{\omega -k}%
\right) ,  \label{pil}
\end{equation}

\begin{equation}
\text{\ \ }\pi _{t}^{\left( e\right) }=\frac{3}{2}\omega _{pe}^{2}\left(
1+\left( \frac{\omega ^{2}}{k^{2}}-1\right) \left( 1-\frac{\omega }{2k}\ln 
\frac{\omega +k}{\omega -k}\right) \right) .  \label{pit}
\end{equation}

Here the electron plasma frequency and the Debye screening distance are
defined as 
\begin{equation}
\omega _{pe}^{2}=\frac{4}{3\pi }e^{2}\mu _{e}^{2}=\frac{4\pi n_{e}e^{2}}{\mu
_{e}},\ \ \ \ \ \ \ \ \ \ \ \ \ \frac{1}{D_{e}^{2}}=3\omega _{pe}^{2},
\label{omega}
\end{equation}
with $\mu _{e}$ and $n_{e}$ being, respectively, the chemical potential and
\ the number density of electrons. Our Eq. (\ref{pil}) differs from Eq.
(A39) of the Ref.\cite{BS93} by an extra factor $\left( \omega
^{2}/k^{2}-1\right) $ because our basis $l^{\mu }$, $h^{\mu }$ is different
from that used by Braaten and Segel by the factor $\sqrt{K^{2}}/k$. All
components of the complete tensor (\ref{pi}) identically coincide with that
obtained in \cite{BS93} for the ultrarelativistic case. In the case $\omega >k$%
, interesting for us, when the Landau damping is forbidden, the one-loop
electron polarization functions $\pi _{l,t}^{\left( e\right) }$ are
real-valued.

The real part of polarization functions of the normal proton component is
caused by dynamics of non-paired proton quasi-particles. For $\omega >k$
this contribution is of the order of \cite{LPkin}: 
\begin{equation}
\mathop{\rm Re}%
\pi _{l}^{\left( p\right) }\sim \frac{1}{6D_{p}^{2}}\left( \frac{kV_{F}}{%
\omega }\right) ^{2}\ll \frac{1}{D_{e}^{2}},\,\ \ \ \ \ \ \ \ \ \frac{kV_{F}%
}{\omega }\ll 1,  \label{6}
\end{equation}
where $D_{p}\sim D_{e}$ is the Debye screening distance of protons. This
proton contribution is negligible compared with the contribution of
ultrarelativistic electrons. Thus, we neglect the real part of proton loops
and assume that the real part of the polarization is defined by the electron
contribution (\ref{pil}), (\ref{pit}) only: 
\begin{equation}
\mathop{\rm Re}%
\pi _{l}=\pi _{l}^{\left( e\right) },\ \ \ \ \ \ \ 
\mathop{\rm Re}%
\pi _{t}=\pi _{t}^{\left( e\right) }  \label{7}
\end{equation}

\subsection{Imaginary part of polarizations}

The imaginary part of polarizations, responsible for the in-medium photon
decay, is caused by formation and breaking of the proton Cooper pairs.
\vskip 0.3cm
\psfig{file=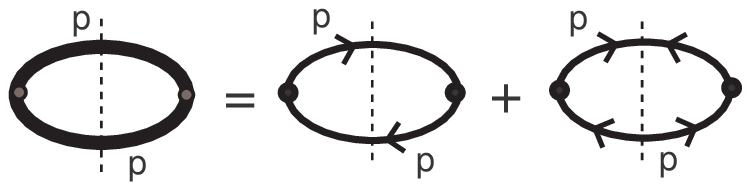}
Fig 1. Feynman graphs contributing to polarization of the proton plasma. 
Two proton loops, in the right-hand side, are
contributions from the normal quasi-particle propagator and from
''so-called'' anomalous propagator which arises in the superconducting phase 
\cite{lp80}.
\vskip 0.3cm
\noindent  The
simplest way to calculate the imaginary part of the polarization tensor
 is an utilizing of the unitarity relation for conversion of a
photon into itself through intermediate state with two quasi-particles. This
corresponds to the loop approximation for the proton polarization tensor, as
shown in Fig. 1.
We consider the retarded polarization tensor. At finite
temperature, the unitarity relation with the intermediate state with two
quasi-particles has to be written as 
\begin{equation}
\frac{2}{\exp \left( \frac{\omega }{T}\right) -1}%
\mathop{\rm Im}%
\pi ^{\mu \nu }=4\pi e^{2}\sum_{i,f}%
\mathop{\rm Tr}%
\left( {\cal M}_{V}^{\mu }{\cal M}_{V}^{\ast \nu }\right) ,  \label{unitV}
\end{equation}
where ${\cal M}_{V}^{\mu }$ is the matrix element of the vector transition
current caused by formation of the Cooper pair, i. e. due to annihilation of
two proton quasi-particles of momentum-spin labels $\left( {\bf p,}s;{\bf p}%
^{\prime },-s\right) $ and the total four-momentum $K$. The summation symbol
includes statistical averaging and summation over initial and final states
of the background, i. e. summation over all initial $p=\left( E,{\bf p}%
\right) $ and final $p^{\prime }=\left( E^{\prime },{\bf p}^{\prime }\right) 
$ states of the proton quasi-particle has to take into account the Pauli
principle via the appropriate blocking factors, with the Fermi distribution
function $f(E)=1/[\exp (E/T)+1]$.

The nonrelativistic limit can be obtained by replacing $\bar{\psi}\gamma
^{0}\psi \rightarrow \hat{\Psi}^{+}\hat{\Psi}$, $\bar{\psi}\gamma _{i}\psi
\rightarrow 0$ , where $\hat{\Psi}$ is the secondary-quantized
nonrelativistic spinor wave function of quasi-protons in superconducting
matter, which is determined by the Bogoliubov transformation. The
transformation for singlet-state pairing is of the form (See e.g., \cite
{lp80}): 
\begin{eqnarray}
\hat{\Psi} &=&\sum_{{\bf p}}\,{\rm \exp }\left( i{\bf pr}\right) \,u_{{\bf p}%
}\,\left( \chi _{s}\hat{\alpha}_{{\bf p,}s}+\chi _{-s}\hat{\alpha}_{{\bf p,-}%
s}\right)  \nonumber \\
&&+\sum_{{\bf p}}{\rm \exp }\left( -i{\bf pr}\right) \,v_{{\bf p}}\,\left(
\chi _{s}\hat{\alpha}_{{\bf p,-}s}^{+}+\chi _{-s}\hat{\alpha}_{{\bf p,}%
s}^{+}\right) ,  \label{wf}
\end{eqnarray}
where ${\bf p}$ is a quasi-particle momentum, and 
\begin{equation}
E=\sqrt{\epsilon ^{2}+\Delta ^{2}}  \label{Ener}
\end{equation}
is its energy with respect to the Fermi level. Near the Fermi surface, at $%
|p-p_{F}|\ll p_{F}$, one has $\epsilon =V_{F}(p-p_{F})$; $s$ enumerates spin
states; $\Delta $ is a superfluid gap at the Fermi surface ($\Delta \ll
p_{F}\,V_{F}$); $V_{F}=p_{F}/M^{\ast }$ is the proton Fermi velocity; $\chi
_{s}$ is a basic spinor ($\chi _{s}^{+}\chi _{s^{\prime }}=\delta
_{ss^{\prime }}$); $\hat{\alpha}_{{\bf p,}s}$ and $\hat{\alpha}_{{\bf p,}%
s}^{+}$ are the annihilation and creation operators, respectively. The
coherence factors are 
\begin{equation}
u_{{\bf p}}=\sqrt{{\frac{1}{2}}\left( 1+{\frac{\epsilon }{E}}\right) },~~~v_{%
{\bf p}}=\sqrt{{\frac{1}{2}}\left( 1-{\frac{\epsilon }{E}}\right) }.
\label{cf}
\end{equation}
The squared matrix element for the proton-pair vector transition has been
calculated by Flowers et al. \cite{FRS76} 
\begin{equation}
\sum_{\sigma }\left| \left\langle {\bf p,}\sigma ;{\bf p}^{\prime },-\sigma
\left| \hat{\Psi}^{+}\hat{\Psi}\right| 0\right\rangle \right| ^{2}=2\left(
u_{{\bf p}}v_{{\bf p}^{\prime }}+u_{{\bf p}^{\prime }}v_{{\bf p}}\right)
^{2}.  \label{8}
\end{equation}
Thus, for non-relativistic protons we have 
\begin{eqnarray}
\sum_{i,f}%
\mathop{\rm Tr}%
\left( {\cal M}_{V}^{\mu }{\cal M}_{V}^{\ast \nu }\right) &=&g^{0\mu
}g^{0\nu }\int {\frac{d^{3}{p}}{(2\pi )^{3}}}\;{\frac{d^{3}{p}^{\prime }}{%
(2\pi )^{3}}}  \nonumber \\
&&(2\pi )^{4}\,\delta ^{(4)}(p+p^{\prime }-K)\;f(E)f(E^{\prime })2\left( u_{%
{\bf p}}v_{{\bf p}^{\prime }}+u_{{\bf p}^{\prime }}v_{{\bf p}}\right) ^{2}.
\label{9}
\end{eqnarray}
The dominant contribution to the integrals comes from proton momenta near
the Fermi surface. As the neutrino-pair momentum is $k\sim T_{c}\ll P_{F}$,
one can neglect ${\bf k}$ in the momentum conservation $\delta $-function,
thus obtaining ${\bf p}^{\prime }=-{\bf p}$. After this simplification, all
integrations become trivial. Assuming $\omega >0$, we obtain 
\begin{equation}
\mathop{\rm Im}%
\pi ^{00}={\,}\frac{8e^{2}\,p_{F}M^{\ast }\Delta ^{2}}{\omega \sqrt{\left(
\omega ^{2}-4\Delta ^{2}\right) }}\tanh \left( \frac{\omega }{4T}\right)
\theta (\omega -2\Delta ),  \label{Impi00}
\end{equation}
where $\theta (x)$ is the Heaviside step-function, $M^{\ast }$ is the
effective proton mass, and 
\begin{equation}
\mathop{\rm Im}%
\pi ^{i0}=%
\mathop{\rm Im}%
\pi ^{0j}=%
\mathop{\rm Im}%
\pi ^{ij}=0  \label{Impiij}
\end{equation}
By the use of decomposition of the polarization tensor 
\begin{equation}
\pi ^{\,\rho \mu }\left( K\right) =\pi _{l}\left( K\right) L^{\rho \mu }+\pi
_{t}\left( K\right) T^{\rho \mu }  \label{10}
\end{equation}
we obtain finally: 
\begin{eqnarray}
\mathop{\rm Im}%
\pi _{l}\left( K\right) &=&-8\frac{k^{2}}{K^{2}}{\,}\frac{%
\,e^{2}p_{F}M^{\ast }\Delta ^{2}}{\omega \sqrt{\left( \omega ^{2}-4\Delta
^{2}\right) }}\tanh \left( \frac{\omega }{4T}\right) \theta (\omega -2\Delta
),  \nonumber \\
\mathop{\rm Im}%
\pi _{t}\left( K\right) &=&0,  \label{Impl}
\end{eqnarray}

\subsection{In-medium photon propagator}

The in-medium photon propagator, taking into account the medium polarization
due to superconducting and normal induced currents, can be found by
summation of all ring polarization diagrams. This can be performed with
the aid of the Dyson equation graphically depicted in Fig. 2.
\vskip 0.3cm
\psfig{file=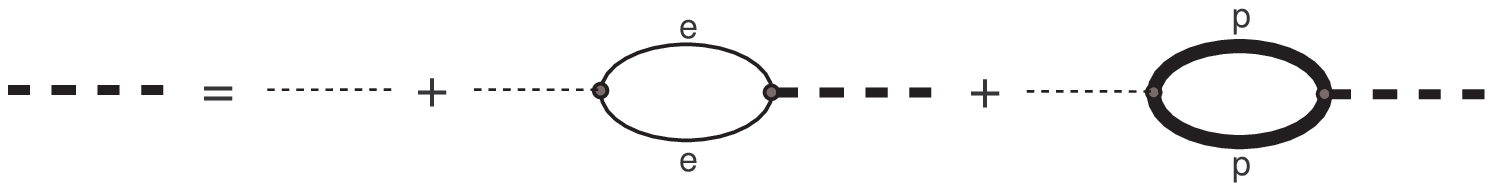}
Fig 2. The graphical representation for the Dyson's equation. The in-medium 
photon, shown by thick dashed line, is the infinite sum of ring 
particle-hole diagrams, connected by thin dashed line of the massive photon 
in the superconductor.  
\vskip 0.3cm
\noindent 
Here the thin dashed line means the above massive photon, which, in the wave
number-frequency space has the propagator of the standard form 
\begin{equation}
D_{0}^{\mu \nu }\left( K\right) =\frac{4\pi }{K^{2}-m_{\gamma }^{2}}\left(
g^{\mu \nu }-\frac{K^{\mu }K^{\nu }}{m_{\gamma }^{2}}\right) .  \label{D0}
\end{equation}
An analytical form of this equation is 
\begin{equation}
D^{\mu \nu }\left( K\right) =D_{0}^{\mu \nu }\left( K\right) +\frac{1}{4\pi }%
D_{0}^{\mu \lambda }\left( K\right) \pi _{\lambda \rho }\left( K\right)
D^{\rho \nu }\left( K\right).  \label{dys}
\end{equation}
Solution is: 
\begin{equation}
D^{\mu \nu }\left( K\right) =D_{l}\left( K\right) L^{\mu \nu }+D_{t}\left(
K\right) T^{\mu \nu }-\frac{4\pi }{m_{\gamma }^{2}}\frac{K^{\mu }K^{\mu }}{%
K^{2}},  \label{D}
\end{equation}
where the longitudinal photon propagators is: 
\begin{equation}
D_{l}\left( K\right) =\frac{4\pi }{K^{2}-m_{\gamma }^{2}-%
\mathop{\rm Re}%
\pi _{l}-i%
\mathop{\rm Im}%
\pi _{l}}.  \label{Dl}
\end{equation}
For $\omega >k$, the imaginary part of the transverse polarization vanishes,
therefore the retarded transverse propagator has the form:

\begin{equation}
D_{t}\left( K\right) =\frac{4\pi }{K^{2}-m_{\gamma }^{2}-%
\mathop{\rm Re}%
\pi _{t}+i0}  \label{Dt}
\end{equation}

Due to conservation and gauge invariance of the vector current $\left(
K_{\lambda }\pi ^{\,\lambda \mu }=\pi ^{\,\lambda \mu }K_{\mu }=0\right) $,
the last term of Eq. (\ref{D}) can be omitted.

According to Eqs. (\ref{Dl}), (\ref{Dt}), the longitudinal 
and transverse electromagnetic field obey the following 
equations.
\begin{equation}
\left( -\omega ^{2}+k^{2}+\pi _{l}\left( \omega ,k\right) +m_{\gamma
}^{2}\right) \Phi \left( \omega ,k\right) =0,  \label{Scal}
\end{equation}
\begin{equation}
\left( -\omega ^{2}+k^{2}+\pi _{t}\left( \omega ,k\right) +m_{\gamma
}^{2}\right) {\bf A}\left( \omega ,k\right) =0.  \label{Vect}
\end{equation}
As $%
\mathop{\rm Re}%
\pi _{l,t}\sim \omega _{pe}^{2}\gg m_{\gamma }^{2}$, the high-frequency
polarization of the medium is mostly caused by the electron gas, and the
superconducting condensate of protons makes only minor correction to the
eigen photon modes in the medium. The longitudinal eigen mode of
oscillations has, however, a damping because of absorption of the photon 
by proton Cooper pairs.

One can clarify the physical nature of the photon mass by considering a
static limit. In this case, presence of the superconducting condensate is
crucial for the vector electromagnetic field. As Re$\pi _{t}\left(
0,k\right) =0$, Eq. (\ref{Vect}) has the following static form 
\begin{equation}
\left( k^{2}+m_{\gamma }^{2}\right) {\bf A}\left( 0,k\right) =0.  \label{11}
\end{equation}
This equation has no non-zero solutions for real $k$. Thus, the static
magnetic field in the medium must vanish. This is well known the Meissner
effect. The field 
\begin{equation}
{\bf A}\left( r\right) =\int \frac{d^{3}k}{\left( 2\pi \right) ^{3}}\exp
\left( i{\bf kr}\right) \frac{4\pi {\bf j}_{ext}\left( k\right) }{%
k^{2}+m_{\gamma }^{2}},  \label{12}
\end{equation}
produced in the medium by any external static electric current ${\bf j}%
_{ext}\left( k\right) $, is screened in a distance $r$ as ${\bf \exp }\left(
-m_{\gamma }r\right) .$ Therefore, the photon mass $m_{\gamma }$ can be
identified with the inverse penetration depth of the superconductor.

\subsection{Weak interactions}

For convenience, we write the low-energy Lagrangian of the proton and
electron interaction with a neutrino field in vacuum as follows 
\begin{equation}
{\cal L}_{vac}=\frac{G_{F}}{\sqrt{2}}j_{\left( p,e\right) }^{\mu }j_{\mu }.
\label{H}
\end{equation}
Here $G_{F}$ is the Fermi coupling constant, and the neutrino weak current
is of the standard form 
\begin{equation}
j_{\mu }=\bar{\nu}\gamma _{\mu }\left( 1-\gamma _{5}\right) \nu .  \label{14}
\end{equation}
By these notations, the weak current of a bare proton 
\begin{equation}
j_{p}^{\mu }=\bar{\psi}\gamma ^{\mu }(\tilde{C}_{V}-\tilde{C}_{A}\gamma
_{5})\psi ,  \label{jp}
\end{equation}
where $\psi $ stands for the proton field, is constructed with reduced
constants of the vector $\tilde{C}_{V}=0.5-2\sin ^{2}\theta _{W}\simeq 0.04$
and axial-vector $\tilde{C}_{A}=0.5g_{A}$ coupling; $g_{A}=1.26$, and $%
\theta _{W}$ is the Weinberg angle. Reserving the capital letter notations $%
\tilde{C}_{V}$ and $\tilde{C}_{A}$ \ for proton coupling constants, we will
use, at the same time, small letters, $c_{V},$ for electron coupling
constants with $c_{V}=0.5+2\sin ^{2}\theta _{W}\simeq 0.96$ for electron
neutrinos, and $c_{V}^{\prime }=-0.5+2\sin ^{2}\theta _{W}\simeq -0.04$ \
for muon and tau neutrinos. The electron weak current is then of the form 
\begin{equation}
j_{e}^{\mu }=\bar{u}\gamma ^{\mu }(c_{V}-c_{A}\gamma _{5})u.  \label{15}
\end{equation}

Strictly speaking, the proton weak coupling vertex should be renormalized
due to nucleon-nucleon correlations in nuclear matter. However, in vertices
with neutral currents the nucleon correlations are weakened \cite{52}, and
are negligible with respect to the plasma effects under consideration (See
Appendix A). Therefore we will neglect the nuclear renormalization of the
weak nucleon vertex. As for the electromagnetic vertex of the proton, its
renormalization is also negligible for the matter density of interest \cite
{Mig}

\section{Neutrino emissivity}

\label{sec:emis}

We consider the total energy which is emitted into neutrino pairs per unit
volume and time. By the Optical theorem, for one neutrino flavor, the
emissivity is given by the following formula: 
\begin{equation}
Q=\frac{G_{F}^{2}}{2}\int \;\frac{\omega }{\exp \left( \frac{\omega }{T}%
\right) -1}%
2\mathop{\rm Im}%
\tilde{\Pi}^{\mu \nu }\left( K\right) 
\mathop{\rm Tr}%
\left( j_{\mu }j_{\nu }^{\ast }\right) \frac{d^{3}k_{1}}{2\omega _{1}(2\pi
)^{3}}\frac{d^{3}k_{2}}{2\omega _{2}(2\pi )^{3}},  \label{rate1}
\end{equation}
where the integration goes over the phase volume of neutrino and
antineutrino of the total energy $\omega =\omega _{1}+\omega _{2}$ and the
total momentum ${\bf k=k}_{1}+{\bf k}_{2}$. In this formula $\tilde{\Pi}%
^{\mu \nu }$ is the retarded polarization tensor of the medium which has
ends at the weak vertex $\left( \tilde{C}_{V}\gamma _{\mu }-\tilde{C}%
_{A}\gamma _{\mu }\gamma _{5}\right) $. In the absence of external magnetic
fields, the parity-violating mixed axial-vector polarization does not
contribute to the rate of neutrino-pair production. In fact, by inserting $%
\int d^{4}K\delta ^{\left( 4\right) }\left( K-k_{1}-k_{2}\right) =1$ in this
equation, and making use of the Lenard's integral \ 
\begin{eqnarray}
&&\int \frac{d^{3}k_{1}}{2\omega _{1}}\frac{d^{3}k_{2}}{2\omega _{2}}\delta
^{\left( 4\right) }\left( K-k_{1}-k_{2}\right) 
\mathop{\rm Tr}%
\left( j^{\mu }j^{\nu \ast }\right)  \nonumber \\
&=&\frac{4\pi }{3}\left( K_{\mu }K_{\nu }-K^{2}g_{\mu \nu }\right) \theta
\left( K^{2}\right) \theta \left( \omega \right) ,  \label{16}
\end{eqnarray}
where $\theta (x)$ is the Heaviside step function, we can write 
\begin{equation}
Q=\frac{G_{F}^{2}}{48\pi ^{5}}\int \;\frac{\omega }{\exp \left( \frac{\omega 
}{T}\right) -1}%
\mathop{\rm Im}%
\tilde{\Pi}^{\mu \nu }\left( K\right) \left( K_{\mu }K_{\nu }-K^{2}g_{\mu
\nu }\right) \theta \left( K^{2}\right) \theta \left( \omega \right) d\omega
d^{3}k.  \label{QQ}
\end{equation}
Since the axial-vector polarization has to be an antisymmetric tensor, its
contraction in (\ref{QQ}) with the symmetric tensor $K_{\mu }K_{\nu
}-K^{2}g_{\mu \nu }$ vanishes. Thus, we can take the weak polarization tensor 
as the sum of vector-vector and axial-axial pieces.

\section{Loop approximation}

\label{sec:loop}

To simplest approximation one can write 
\begin{equation}
\tilde{\Pi}_{{\rm loop}}^{\mu \nu }=\tilde{C}_{V}^{2}\Pi _{V}^{\mu \nu }+%
\tilde{C}_{A}^{2}\Pi _{A}^{\mu \nu },  \label{PiLoop}
\end{equation}
where the polarization tensors $\Pi _{V,A}^{\mu \nu }$ are given by the same
diagrams of Fig. 1, but with the ends at weak vertex. Namely, the
vector-vector tensor $\Pi _{V}^{\mu \nu }$ has ends at the vector weak
vertex, while the axial-axial tensor $\Pi _{A}^{\mu \nu }$ begins and ends
at axial vertex.

Obviously, the imaginary part of the weak polarization tensor can be
obtained in the same way as the imaginary part of electromagnetic
polarizations. The electromagnetic polarization differs from the weak
vector-vector polarization only by a constant factor $4\pi e^{2}$, which is
included into the definition of the electromagnetic tensor. Thus, 
\begin{equation}
\mathop{\rm Im}%
\Pi _{V}^{\mu \nu }=\frac{1}{4\pi e^{2}}%
\mathop{\rm Im}%
\pi ^{\mu \nu }.  \label{piPi}
\end{equation}
By the use of decomposition of the vector weak polarization
tensor 
\begin{equation}
\Pi _{V}^{\,\rho \mu }\left( K\right) =\Pi _{l}\left( K\right) L^{\rho \mu
}+\Pi _{t}\left( K\right) T^{\rho \mu }  \label{decomp}
\end{equation}
we obtain

\begin{eqnarray}
\mathop{\rm Im}%
\Pi _{l}\left( K\right) &=&-\frac{2}{\pi }\frac{k^{2}}{K^{2}}{\,}\frac{%
\,p_{F}M^{\ast }\Delta ^{2}}{\omega \sqrt{\left( \omega ^{2}-4\Delta
^{2}\right) }}\tanh \left( \frac{\omega }{4T}\right) \theta (\omega -2\Delta
),  \nonumber \\
\mathop{\rm Im}%
\Pi _{t}\left( K\right) &=&0,  \label{ImPl}
\end{eqnarray}

As mentioned above, in the case of non-relativistic protons, the total spin
of the Cooper pair in the $^{1}S_{0}$-state is zero, therefore the axial
contribution is proportional to a small parameter $V_{F}^{2}$. However, due
to numerical smallness of the proton weak vector coupling constant the axial
contribution should be taken into account. Considering the non-relativistic
axial-vector current of the proton to accuracy $V_{F}$ we have to include
also the relativistic correction to its temporal component by replacing $%
\bar{\psi}\gamma ^{0}\gamma ^{5}\psi \rightarrow \hat{\Psi}^{+}\hat{\varphi}+%
\hat{\varphi}^{+}\hat{\Psi}$, $\bar{\psi}\gamma _{i}\gamma ^{5}\psi
\rightarrow \hat{\Psi}^{+}{\sigma }_{i}\hat{\Psi}$. Here the proton bispinor 
$\psi $ is presented as combination of an upper spinor $\hat{\Psi}$ and
lower spinor $\hat{\varphi}=-i{\sigma}_{i}{\nabla}_{i} \hat{\Psi}/(2M)$, 
where $ M $ is the bare proton mass, and ${\sigma}_{i}$ are the Pauli matrices.

Direct calculation shows that the tensor $I_{A}^{\mu \nu }\equiv
\sum_{\sigma }\left( {\cal M}_{A}^{\mu }{\cal M}_{A}^{\ast \nu }\right) $ is
diagonal, with two nontrivial elements \cite{KHY99}: 
\begin{eqnarray}
I_{A}^{00} &=&(u^{\prime }v+v^{\prime }u)^{2}\frac{\left| 
{\bf p-p}^{\prime }\right| ^{2}}{2M^{2}},  \nonumber \\
I_{A}^{11} &=&I_{A}^{22}=I_{A}^{33}=2(u^{\prime
}v-v^{\prime }u)^{2}.  \label{II}
\end{eqnarray}
The subsequent calculations with the help of the formula 
\begin{equation}
\frac{2}{\exp \left( \frac{\omega }{T}\right) -1}%
\mathop{\rm Im}%
\Pi _{A}^{\mu \nu }=\sum_{i,f}%
\mathop{\rm Tr}%
\left( {\cal M}_{A}^{\mu }{\cal M}_{A}^{\ast \nu }\right) ,  \label{17}
\end{equation}
yield 
\begin{equation}
\mathop{\rm Im}%
\Pi _{A}^{00}=\frac{2}{\pi }{\,}\frac{M^{\ast 2}}{M^{2}}V_{F}^{2}\frac{%
\,p_{F}M^{\ast }\Delta ^{2}}{\omega \sqrt{\left( \omega ^{2}-4\Delta
^{2}\right) }}\tanh \left( \frac{\omega }{4T}\right) \theta (\omega -2\Delta
)  \label{ImPiA00}
\end{equation}
\begin{equation}
\mathop{\rm Im}%
\Pi _{A}^{11}=%
\mathop{\rm Im}%
\Pi _{A}^{22}=%
\mathop{\rm Im}%
\Pi _{A}^{33}=\frac{1}{3\pi }k^{2}\allowbreak V_{F}^{2}\frac{p_{F}M^{\ast
}\Delta ^{2}}{\omega ^{3}\sqrt{\omega ^{2}-4\Delta ^{2}}}\tanh \left( \frac{%
\omega }{4T}\right) \theta (\omega -2\Delta )  \label{ImPiA11}
\end{equation}

Once the imaginary part of the polarization is calculated, one can evaluate
the $\nu \bar{\nu}$ emissivity as indicated in Eq. (\ref{QQ}). By the use of
the Eqs. (\ref{PiLoop}), (\ref{decomp}), (\ref{ImPl}), (\ref{ImPiA00}), (\ref
{ImPiA11})) we obtain 
\begin{equation}
Q^{{\rm loop}}=Q_{V}^{{\rm loop}}+Q_{A}^{{\rm loop}},  \label{QtotLA}
\end{equation}
where the vector weak coupling contribution is 
\begin{equation}
Q_{V}^{{\rm loop}}=\frac{16}{15\pi ^{5}}{\cal N}_{\nu }G_{F}^{2}\tilde{C}%
_{V}^{2}p_{F}M^{\ast }\allowbreak \Delta ^{7}\int_{1}^{\infty }\frac{z^{5}}{%
\left( \exp \left( yz\right) +1\right) ^{2}\sqrt{\left( z^{2}-1\right) }}%
\,dz,  \label{QVloop}
\end{equation}
$\allowbreak $with $y=\Delta /T$ and ${\cal N}_{\nu }$ being the number of
neutrino flavors. The axial-vector coupling gives: 
\begin{equation}
Q_{A}^{{\rm loop}}=\left( \frac{M^{\ast 2}}{M^{2}}+\frac{11}{42}\right)
V_{F}^{2}\,\frac{\tilde{C}_{A}^{2}}{\tilde{C}_{V}^{2}}Q_{V}^{{\rm loop}}
\label{QA}
\end{equation}
\ This result reproduces the emissivity obtained in \cite{FRS76} and \cite
{KHY99}. In spite of the fact that the Fermi velocity of protons is small,
the axial term (\ref{QA}) is typically larger, than the vector term 
(\ref{QVloop}) due to $\tilde{C}_{A}^{2}/\tilde{C}_{V}^{2}\gg 1$.

\section{Random-phase approximation}

\label{sec:RPA}

\subsection{RPA polarizations}

In the above loop approximation we ignored electromagnetic correlations
among the charged particles in a QED plasma. To incorporate the collective
plasma effects, one can use the Random phase approximation (RPA). In the case
under consideration, the RPA means a summation of all ring diagrams
connected by the line of a massive photon. One can perform the above
summation with the aid of the in-medium photon propagator $D_{\mu \nu
}\left( K\right) $, which is solution of the Dyson equation collecting all
ring particle-hole diagrams, responsible for the electromagnetic
polarization of the medium, as shown in Fig. 2. In this way, the graphical
representation of the RPA weak polarization tensor $\tilde{\Pi}_{V}^{\mu \nu
}$ is of the form shown in Fig. 3.
\vskip0.3cm \psfig{file=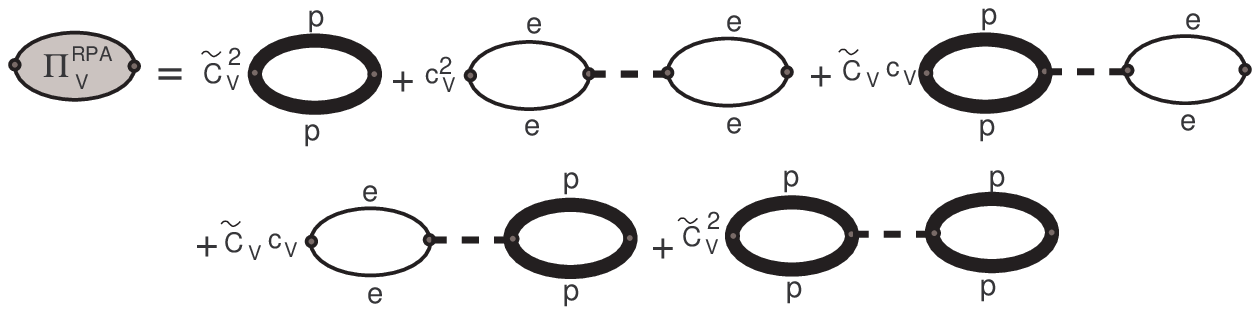} Fig 3. The graphical representation for the 
RPA polarization tensor. The in-medium 
photon, shown by thick dashed line, is a solution of the Dyson equation 
collecting the infinite sum of ring 
particle-hole diagrams, connected by the line of the massive photon 
in superconductor. \vskip%
0.3cm 
\noindent Here the first term describes the
contribution, which comes directly from the pairing protons. The other terms
are caused by the plasma polarization. Analytically one has: 
\begin{eqnarray}
\mathop{\rm Im}%
\tilde{\Pi}_{l}^{{\rm RPA}} &=&\frac{1}{4\pi e^{2}}\left[ \tilde{C}_{V}^{2}%
\mathop{\rm Im}%
\pi _{l}+\frac{1}{4\pi }\left( c_{V}^{2}%
\mathop{\rm Im}%
D_{l}\left( 
\mathop{\rm Re}%
\pi _{l}\right) ^{2}\right. \right.  \nonumber \\
&&\left. \left. -2\tilde{C}_{V}c_{V}%
\mathop{\rm Im}%
\pi _{l}%
\mathop{\rm Re}%
D_{l}%
\mathop{\rm Re}%
\pi _{l}-\tilde{C}_{V}^{2}\left( 
\mathop{\rm Im}%
\pi _{l}\right) ^{2}%
\mathop{\rm Im}%
D_{l}\right) \right] .  \label{RPA}
\end{eqnarray}
Here we took into account the fact that 
\begin{equation}
\mathop{\rm Re}%
\pi _{l}^{\left( p\right) }=%
\mathop{\rm Im}%
\pi _{l}^{\left( e\right) }=0  \label{18}
\end{equation}
as discussed above. An extra factor $\left( 4\pi e^{2}\right) ^{-1}$ appears
in Eq. (\ref{RPA}) because we replaced weak polarization tensors by
electromagnetic tensors of the plasma. One more factor $\left( 4\pi
\right) ^{-1}$ appears because the factor $4\pi $ is traditionally included
into the definition of electromagnetic polarization tensors. The minus sign,
in front of terms, which are proportional to $\tilde{C}_{V}c_{V}$, is due to
the fact that those diagrams include the proton charge $e>0$ at one of 
electromagnetic vertices of the virtual photon, and the electron charge $-e$
at the opposite end. The corresponding factor $e^{2}$ which appears
accordingly to the diagrams with a photon line, is also included into the
definition of polarization tensors. For the transverse weak polarization we
obtain 
\begin{equation}
\mathop{\rm Im}%
\tilde{\Pi}_{t}^{{\rm RPA}}=\frac{c_{V}^{2}}{8\pi ^{2}e^{2}}%
\mathop{\rm Im}%
D_{t}\left( 
\mathop{\rm Re}%
\pi _{t}\right) ^{2}.  \label{19}
\end{equation}
By replacing the photon propagators by their explicit expressions (\ref{Dl}%
), (\ref{Dt}) we obtain: 
\begin{eqnarray}
\mathop{\rm Im}%
\tilde{\Pi}_{l}^{{\rm RPA}} &=&\frac{%
\mathop{\rm Im}%
\pi _{l}}{4\pi e^{2}}\left[ \frac{\tilde{C}_{V}^{2}\left( K^{2}-m_{\gamma
}^{2}-%
\mathop{\rm Re}%
\pi _{l}\right) ^{2}}{\left( K^{2}-m_{\gamma }^{2}-%
\mathop{\rm Re}%
\pi _{l}\right) ^{2}+\left( 
\mathop{\rm Im}%
\pi _{l}\right) ^{2}}\right.  \nonumber \\
&&\left. -\frac{2\tilde{C}_{V}c_{V}\left( K^{2}-m_{\gamma }^{2}-%
\mathop{\rm Re}%
\pi _{l}\right) 
\mathop{\rm Re}%
\pi _{l}}{\left( K^{2}-m_{\gamma }^{2}-%
\mathop{\rm Re}%
\pi _{l}\right) ^{2}+\left( 
\mathop{\rm Im}%
\pi _{l}\right) ^{2}}+\frac{c_{V}^{2}\left( 
\mathop{\rm Re}%
\pi _{l}\right) ^{2}}{\left( K^{2}-m_{\gamma }^{2}-%
\mathop{\rm Re}%
\pi _{l}\right) ^{2}+\left( 
\mathop{\rm Im}%
\pi _{l}\right) ^{2}}\right] .  \label{PiTilda}
\end{eqnarray}
\begin{equation}
\mathop{\rm Im}%
\tilde{\Pi}_{t}^{{\rm RPA}}=\frac{c_{V}^{2}}{4\pi e^{2}}\left( 
\mathop{\rm Re}%
\pi _{t}\right) ^{2}\delta \left( K^{2}-m_{\gamma }^{2}-%
\mathop{\rm Re}%
\pi _{t}\right)  \label{21}
\end{equation}

The imaginary part of the transverse polarization corresponds to the
transverse eigen photon mode, which has the dispersion defined by Eq. (\ref
{Vect}). The eigen-mode frequency $\omega _{t}\left( k\right) $ \ is larger
than the plasma frequency of electrons. Thus, the term $%
\mathop{\rm Im}%
\tilde{\Pi}_{t}$ describes the contribution from the decay of real
transverse photon of energy $\omega _{t}\left( k\right) >\omega _{pe}\gg
T_{c}$. At temperatures $T<T_{c}\ll \omega _{pe}$, the number of such
photons in the medium is exponentially suppressed, therefore this term can
be neglected.

By the same reason we can neglect the RPA corrections to the axial medium
polarization. Detailed consideration shows that these corrections occur due
to exchange of transverse photons between charged particles in the plasma.
As the imaginary part of the transverse photon is the delta-function, the
RPA corrections to the axial polarization are of the form 
\begin{equation}
\delta 
\mathop{\rm Im}%
\tilde{\Pi}_{t}^{{\rm RPA}}=\frac{c_{V}^{2}}{4\pi e^{2}}\left( 
\mathop{\rm Re}%
\pi _{AV}\right) ^{2}\delta \left( K^{2}-m_{\gamma }^{2}-%
\mathop{\rm Re}%
\pi _{t}\right) ,  \label{delta}
\end{equation}
where $%
\mathop{\rm Re}%
\pi _{AV}$ is the mixed axial-vector polarization function of the electron
plasma. Obviously such a correction is caused by the transverse plasmon
decay. Thus, we take 
\begin{equation}
\left( 
\mathop{\rm Im}%
\tilde{\Pi}_{A}^{\mu \nu }\right) ^{{\rm RPA}}=\left( 
\mathop{\rm Im}%
\tilde{\Pi}_{A}^{\mu \nu }\right) ^{{\rm loop}}  \label{ImARPA}
\end{equation}

Further, as the polarization function $\pi _{l}\sim \omega _{pe}^{2}$, we
can neglect $K^{2}\sim T_{c}^{2}\ll $ $\omega _{pe}^{2}$ in 
Eq. (\ref{PiTilda}). By replacing also the common factor as specified by
Eq. (\ref{piPi}) we obtain: 
\begin{eqnarray}
\mathop{\rm Im}%
\tilde{\Pi}_{l}^{{\rm RPA}} &=&%
\mathop{\rm Im}%
\Pi _{l}\left[ \tilde{C}_{V}^{2}\frac{\left( m_{\gamma }^{2}+%
\mathop{\rm Re}%
\pi _{l}\right) ^{2}}{\left( m_{\gamma }^{2}+%
\mathop{\rm Re}%
\pi _{l}\right) ^{2}+\left( 
\mathop{\rm Im}%
\pi _{l}\right) ^{2}}\right.  \nonumber \\
&&\left. +2c_{V}\tilde{C}_{V}\frac{\left( m_{\gamma }^{2}+%
\mathop{\rm Re}%
\pi _{l}\right) 
\mathop{\rm Re}%
\pi _{l}}{\left( m_{\gamma }^{2}+%
\mathop{\rm Re}%
\pi _{l}\right) ^{2}+\left( 
\mathop{\rm Im}%
\pi _{l}\right) ^{2}}+c_{V}^{2}\frac{\left( 
\mathop{\rm Re}%
\pi _{l}\right) ^{2}}{\left( m_{\gamma }^{2}+%
\mathop{\rm Re}%
\pi _{l}\right) ^{2}+\left( 
\mathop{\rm Im}%
\pi _{l}\right) ^{2}}\right] ,  \label{PiRPA}
\end{eqnarray}
where the loop polarizations are defined in Eqs. (\ref{ImPl}), (\ref{Impl}),
and (\ref{pil}); and $m_{\gamma }$, given by Eq.(\ref{mg}), is the mass
acquired by a photon in the superconductor due to the Higgs effect. As
noticed above, this mass can be identified with the inverse penetration
depth of the superconductor. If to eliminate the plasma polarization by
taking formally $\pi _{l}\rightarrow 0$ then Eq. (\ref{PiRPA}) reproduces
the result obtained in the loop approximation. However, the plasma
polarization strongly modifies that result.

The first term in the brackets of Eq. (\ref{PiRPA}) is the summarized
contribution of two diagrams of the Fig. 3, which are proportional to $%
\tilde{C}_{V}^{2}$. This term is always less than unity. This means, that
ambient protons screen the weak vector interaction of the pairing protons
with a neutrino field.

Let us estimate relative contributions of different terms in Eq. (\ref{PiRPA}%
). According to Eqs. (\ref{mg}) and (\ref{omega}), the penetration depth of
the superconductor is typically larger than the electron Debye screening
distance, i.e. $m_{\gamma }^{2}\ll D_{e}^{-2}$. Thus, $m_{\gamma
}^{2}\lesssim \left| 
\mathop{\rm Re}%
\pi _{l}\right| $, and therefore $\left| m_{\gamma }^{2}+%
\mathop{\rm Re}%
\pi _{l}\right| \sim \left| 
\mathop{\rm Re}%
\pi _{l}\right| $.

For electron neutrinos one has $\tilde{C}_{V}^{2}/c_{V}^{2}\simeq
1.\,\allowbreak 74\times 10^{-3}$ and $\tilde{C}_{V}/c_{V}$:$\simeq $ $%
4.\,\allowbreak 1\,7\times 10^{-2}$. Thus, for electron neutrinos, the first
two terms in Eq. (\ref{PiRPA}) are small compared with the term, which is
proportional to $c_{V}^{2}$. For muon and tauon neutrinos all the terms are
about $10^{-3}$ of the leading contribution for electron neutrinos.
Numerical tests have shown that, to accuracy less than one percent, the
vector contribution to the emissivity can be described by the leading term
which is proportional to $c_{V}^{2}$. Notice that relativistic corrections,
which we do not take into account, are of order $V_{F}^{2}$ ($%
V_{F}^{2}\simeq 0.06$ for a $\beta $-equilibrium nuclear matter of the
density $\rho =2\rho _{0})$. To this accuracy, the above small terms must be
ignored. Thus we obtain the following analytic expression for the imaginary
part of the vector-vector RPA tensor 
\begin{equation}
\mathop{\rm Im}%
\tilde{\Pi}_{l}^{{\rm RPA}}\simeq c_{V}^{2}%
\mathop{\rm Im}%
\Pi _{l}\frac{\left( 
\mathop{\rm Re}%
\pi _{l}\right) ^{2}}{\left( m_{\gamma }^{2}+%
\mathop{\rm Re}%
\pi _{l}\right) ^{2}+\left( 
\mathop{\rm Im}%
\pi _{l}\right) ^{2}}.  \label{PiApprox}
\end{equation}
This approximation corresponds to the diagram with two electron loops in
Fig. 3.

\subsection{Including muons}

When the chemical potential of electrons $\mu _{e}$ is larger than the muon
mass $m_{\mu }$, the $\beta $-equilibrium nuclear matter contains 
muons. Generalization of Eq. (\ref{PiApprox}) to the case of p, e, $\mu $
plasma is obvious. One has to add the corresponding diagrams with muon loops
both to the Dyson's equation for the photon propagator and to the RPA
polarization tensor. This yields 
\begin{equation}
\sum_{\nu _{e},\nu _{\mu },\nu _{\tau }}%
\mathop{\rm Im}%
\tilde{\Pi}_{l}^{{\rm RPA}}\simeq c_{V}^{2}%
\mathop{\rm Im}%
\Pi _{l}\left[ \frac{\left( 
\mathop{\rm Re}%
\pi _{l}^{\left( e\right) }\right) ^{2}+2\left( 
\mathop{\rm Re}%
\pi _{l}^{\left( \mu \right) }\right) ^{2}}{\left( m_{\gamma }^{2}+%
\mathop{\rm Re}%
\pi _{l}^{\left( e\right) }+%
\mathop{\rm Re}%
\pi _{l}^{\left( \mu \right) }\right) ^{2}+\left( 
\mathop{\rm Im}%
\pi _{l}\right) ^{2}}\right] ,  \label{PiMuRPA}
\end{equation}
where the muonic polarization function is 
\begin{equation}
\pi _{l}^{\left( \mu \right) }=\frac{1}{D_{\mu }^{2}}\left( 1-\frac{\omega
^{2}}{k^{2}}\right) \left( 1-\frac{\omega }{2kv_{F}}\ln \frac{\omega +kv_{F}%
}{\omega -kv_{F}}\right).   \label{pilmu}
\end{equation}
In Eq. (\ref{PiMuRPA}), the fact is taken into account that the weak vector
coupling constant for a muon radiating muonic, and tauon neutrinos is the
same as that for an electron radiating electron neutrinos. By this reason
the muonic polarization function comes in the numerator of Eq. (\ref{PiMuRPA}) 
with the factor of 2.  The muonic Fermi velocity 
\begin{equation}
v_{F}=\frac{\sqrt{\mu _{\mu }^{2}-m_{\mu }^{2}}}{\mu _{\mu }},  \label{vmu}
\end{equation}
and the Debye screening distance 
\begin{equation}
D_{\mu }^{-2}=3\frac{4\pi n_{\mu }e^{2}}{\mu _{\mu }v_{F\mu }^{2}}=\frac{4}{%
\pi }e^{2}\left( \mu _{\mu }^{2}-m_{\mu }^{2}\right) ^{1/2}\mu _{\mu }.
\label{Dmu}
\end{equation}
depend on the muonic chemical potential $\mu _{\mu }$. 

Numerical evaluation under condition of chemical equilibrium  
$\mu _{\mu}=\mu _{e}$ shows (see Fig. 7) that muons
introduce a minor correction to neutrino emissivity.

\subsection{RPA emissivity}

By inserting the imaginary part of the vector-vector RPA polarization tensor 
\begin{equation}
\mathop{\rm Im}%
\left( \tilde{\Pi}_{V}^{{\rm RPA}}\right) ^{\mu \nu }=%
\mathop{\rm Im}%
\tilde{\Pi}_{l}^{{\rm RPA}}L^{\mu \nu }  \label{22}
\end{equation}
into the emissivity equation 
\begin{equation}
Q_{V}^{{\rm RPA}}=\frac{G_{F}^{2}}{48\pi ^{5}}\int \;\frac{\omega }{\exp
\left( \frac{\omega }{T}\right) -1}%
\mathop{\rm Im}%
\left( \tilde{\Pi}_{V}^{{\rm RPA}}\right) ^{\mu \nu }\left( K_{\mu }K_{\nu
}-K^{2}g_{\mu \nu }\right) \theta \left( K^{2}\right) \theta \left( \omega
\right) d\omega d^{3}k,  \label{23}
\end{equation}
we obtain the following vector contribution to emissivity: 
\begin{equation}
Q_{V}^{{\rm RPA}}=-\frac{G_{F}^{2}c_{V}^{2}}{12\pi ^{4}}\int_{2\Delta
}^{\infty }d\omega \;\int_{0}^{\omega }dk\,k^{2}\frac{\omega \left( \omega
^{2}-k^{2}\right) }{\exp \left( \frac{\omega }{T}\right) -1}%
\mathop{\rm Im}%
\Pi _{l}\frac{\left( 
\mathop{\rm Re}%
\pi _{l}\right) ^{2}}{\left( m_{\gamma }^{2}+%
\mathop{\rm Re}%
\pi _{l}\right) ^{2}+\left( 
\mathop{\rm Im}%
\pi _{l}\right) ^{2}}.  \label{QVe}
\end{equation}
According to above discussion $Q_{V}^{{\rm RPA}}$, given by this formula, is
the $\nu \bar{\nu}$ emissivity of ambient electrons perturbed by the proton
quantum transition. Therefore it is proportional to the square of the
electron vector weak coupling constant, and vanishes when the proton energy
gap is closed, because $%
\mathop{\rm Im}%
\Pi _{l}\left( \Delta =0\right) =0$.

Replacing the polarization functions by their explicit expressions (\ref
{ImPl}), (\ref{Impl}), and (\ref{pil}) we finally obtain: 
\begin{eqnarray}
Q_{V}^{{\rm RPA}} &=&\frac{16}{15\pi ^{5}}G_{F}^{2}c_{V}^{2}p_{F}M^{\ast
}\Delta ^{7}\int_{1}^{\infty }dz\frac{z^{7}\sqrt{z^{2}-1}}{[\exp (zy)+1]^{2}}%
{\,}  \nonumber \\
&&\int_{0}^{1}dx\frac{5x^{4}\left( 1-x^{2}\right) ^{2}\varphi ^{2}\left(
x\right) }{z^{2}\left( z^{2}-1\right) \left( 1-x^{2}\right) ^{2}\left(
\varphi \left( x\right) -\lambda x^{2}\right) ^{2}+x^{8}{\,}\tanh ^{2}\left( 
z y/2 \right) \eta ^{2}}  \label{QV1}
\end{eqnarray}
with 
\begin{equation}
\varphi \left( x\right) =\left( 1-x^{2}\right) \left( 1-\frac{1}{2x}\ln 
\frac{1+x}{1-x}\right) .  \label{24}
\end{equation}
Additionally to the temperature and the energy gap, the emissivity caused by 
the vector weak coupling depends substantially on
two new parameters which does not appear in the loop approximation. The
parameter 
\begin{equation}
\lambda =\ m_{\gamma }^{2}D_{e}^{2}=\frac{\mu _{e}}{3M^{\ast }}\frac{n_{p}}{%
n_{e}}\frac{N_{p}\left( T/T_{c}\right) }{n_{p}}  \label{25}
\end{equation}
is the squared ratio of the electron Debye screening distance to the
penetration depth of the superconductor. The latter equality follows from (%
\ref{omega}) and (\ref{mg}). Here $n_{e}$ is the number density of
electrons; $n_{p}$ is the total proton number density, and $N_{p}\left(
T/T_{c}\right) $ is the number density of paired protons. In the case of
p, e-plasma one has $n_{p}=n_{e}$. Typically $\lambda \sim 0.05\div 0.1$. The
second parameter is $\eta =2\,e^{2}p_{F}M^{\ast }D_{e}^{2}$, which can be
written in the following form 
\begin{equation}
\eta =\frac{\pi }{2}\frac{D_{e}^{2}}{D_{p}^{2}},  \label{eta}
\end{equation}
where 
\begin{equation}
\frac{1}{D_{p}^{2}}=3\,\frac{4\pi e^{2}n_{p}}{MV_{F}^{2}}
\end{equation}
is the inverse Debye screening distance of protons. The parameter $\eta $ is typically about one.

The temperature dependence of the emissivity enters by means of parameter 
\begin{equation}
y=\frac{\Delta \left( T\right) }{T}=\frac{\Delta \left( 0\right) }{T_{c}}%
\frac{\Delta \left( \tau \right) }{\tau \Delta \left( 0\right) }
\end{equation}
with $\tau =T/T_{c}$. For a singlet-state pairing $\Delta \left( 0\right)
/T_{c}=1.\,\allowbreak 76$ (See e.g. \cite{lp80}), therefore the function $y$
depends on the dimensionless temperature $\tau $ only. Thus, the emissivity (%
\ref{QV1}), in [erg s$^{-1}$ cm$^{-3}$], can be written as 
\begin{eqnarray}
Q_{V}^{{\rm RPA}} &=&1.17\times 10^{21}c_{V}^{2}\left( \frac{M^{\ast }}{M}%
\right) ^{2}V_{F}\left( \frac{T_{c}}{10^{9}\ K}\right) ^{7}\,\tau
^{7}y^{7}\int_{1}^{\infty }dz\frac{z^{7}\sqrt{z^{2}-1}}{[\exp (zy)+1]^{2}} 
\nonumber \\
&&\int_{0}^{1}dx\frac{5x^{4}\left( 1-x^{2}\right) ^{2}\varphi ^{2}\left(
x\right) }{z^{2}\left( z^{2}-1\right) \left( 1-x^{2}\right) ^{2}\left(
\varphi \left( x\right) -\lambda x^{2}\right) ^{2}+x^{8}{\,}\tanh ^{2}\left( 
\frac{z}{2}y\right) \eta ^{2}},  \label{QV2}
\end{eqnarray}
where 
\begin{equation}
V_{F}=0.36\frac{M}{M^{\ast }}\left( \frac{n_{p}}{n_{0}}\right) ^{1/3}
\label{27}
\end{equation}
is the proton Fermi velocity; and $n_{0}=0.16\;fm^{-3}$ is the number
density of saturated nuclear matter.

By adding the contribution (\ref{QA}) caused by the axial-vector coupling we
finally obtain the following total emissivity: 
\begin{equation}
Q=Q_{V}^{{\rm RPA}}+Q_{A}  \label{QtotRPA}
\end{equation}
with $Q_{V}^{{\rm RPA}}$ given by Eq. (\ref{QV2}) and 
\begin{eqnarray}
Q_{A} &=&\,1.17\times 10^{21}\tilde{C}_{A}^{2}{\cal N}_{\nu
}\left( \frac{M^{\ast }}{M}\right) ^{2}\left( \frac{M^{\ast 2}}{M^{2}}+\frac{%
11}{42}\right) V_{F}^{3}\left( \frac{T_{c}}{10^{9}\ K}\right) ^{7}\times 
\nonumber \\
&&\tau ^{7}y^{7}\int_{1}^{\infty }\frac{z^{5}}{[\exp (zy)+1]^{2}\sqrt{\left(
z^{2}-1\right) }}\,dz  \label{QQA}
\end{eqnarray}

\section{Numerical estimate}

\label{sec:numer}

To evaluate emissivities (\ref{QV2}), (\ref{QQA}) numerically we use the analytic fit
of the energy gap suggested in \cite{YKL98}: 
\begin{equation}
y\left( \tau \right) =\sqrt{1-\tau }\left( 1.456-\frac{0.157}{\sqrt{\tau }}+%
\frac{1.764}{\tau }\right) ,  \label{vfit}
\end{equation}

The number density of paired protons $N_{p}$ depends on the dimensionless
temperature $\tau $. By the use of the fitted temperature dependence for the
energy gap (\ref{vfit}) one can calculate $N_{p}\left( \tau \right) $ as
follows \cite{lp80}: 
\begin{equation}
\frac{N_{p}\left( \tau \right) }{n_{p}}=1+2\int \frac{df\left( E\right) }{dE}%
d\epsilon ,  \label{EqNp}
\end{equation}
where $E=\sqrt{\epsilon ^{2}+\Delta ^{2}}$, and $f\left( E\right) =1/\left(
\exp \left( E/T\right) +1\right) $.

$\allowbreak \allowbreak \allowbreak \allowbreak \allowbreak \allowbreak
\allowbreak $Thus, we obtain: 
\begin{equation}
\lambda =\frac{\mu _{e}}{3M^{\ast }}\frac{N_{p}\left( \tau \right) }{n_{p}}
\label{psi}
\end{equation}
The function $N_{p}\left( \tau \right) /n_{p}$ is shown in Fig. 4. 
\vskip 0.3cm
\psfig{file=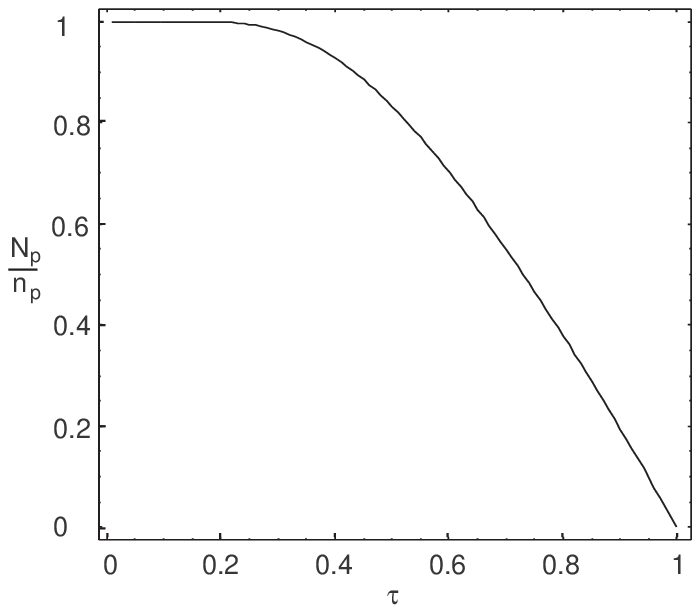}
Fig 4. Relative fraction of paired protons $N_{p}/n_{p}$ versus the
dimensionless temperature $\tau =T/T_{c}$ . \vskip 0.3cm
\noindent
For illustration we consider
matter with total baryon density $\rho =2\rho _{0}$ consisting on neutrons
with a small fraction of protons and electrons corresponding to $\beta $%
-equilibrium. We set the effective proton mass $M^{\ast }=0.7M^{\ast} $. 
To demonstrate efficiency of the
collective effects, the $Q_{V}^{{\rm RPA}}$, given by Eq. (\ref{QV2}), is
plotted in Fig. 5 versus the dimensionless temperature $\tau =T/T_{c}$
together with that obtained by the loop approximation (LA) without
collective effects. One can see that, the collective effects substantially
enhance, the $\nu \bar{\nu}$ emissivity caused by the vector weak current of
protons, but the temperature dependence becomes more sharp. Due to the collective effects the total 
emissivity sharply increases 
when the temperature becomes below than the critical temperature for proton
pairing. The temperature dependence of total emissivity is shown in Fig. 6.
\vskip0.3cm \psfig{file=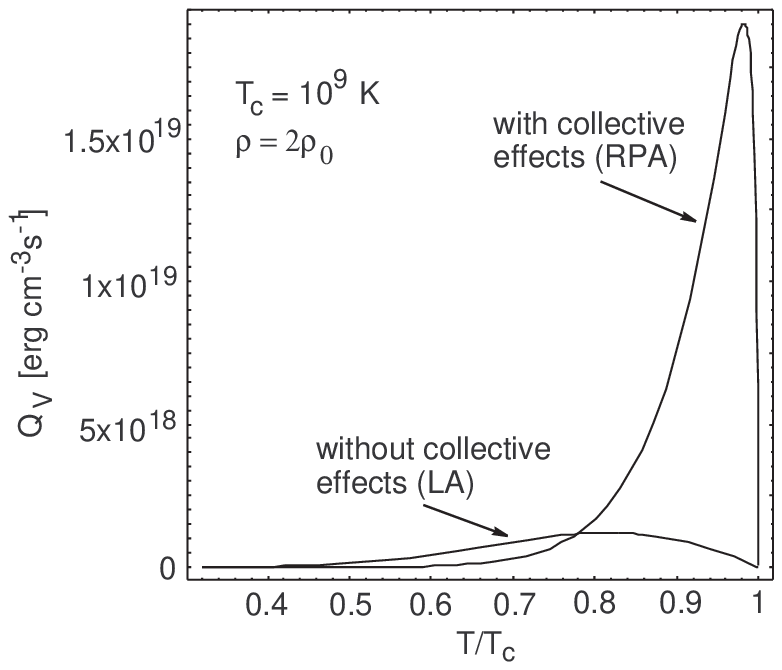} Fig 5. Temperature
dependence of the RPA vector weak current contribution to neutrino
emissivity in comparision with
that obtained in the loop approximation (LA). The emissivities shown versus the dimensionless
temperature $\tau =T/T_{c}$ for $\beta $-equilibrium nuclear matter. \vskip%
0.3cm 

\vskip0.3cm \psfig{file=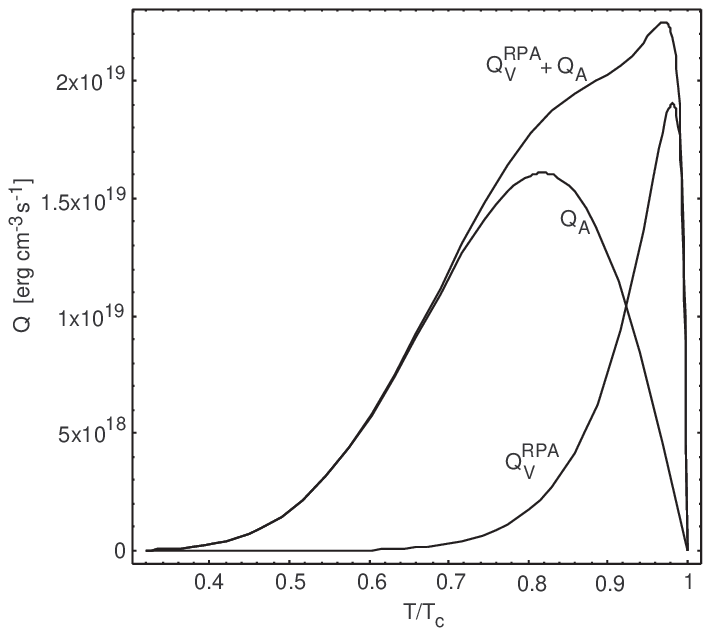} Fig 6. Total emissivity $Q=Q_{V}^{%
{\rm RPA}}+Q_{A}$ versus the dimensionless temperature. 
\vskip0.3cm

As mentioned above, when the chemical potential of electrons is larger than 
the muon mass, the nuclear matter contains muons. In this case the muonic 
contribution to the RPA polarization tensor can be taken into account as 
indicated in Eq. (\ref{PiMuRPA}). The total emissivities calculated with 
and without the muon contribution are plotted in Fig. 7. The calculation was 
made under the condition for chemical equilibrium $\mu _{\mu }=\mu _{e}$. 
	
\vskip0.3cm \psfig{file=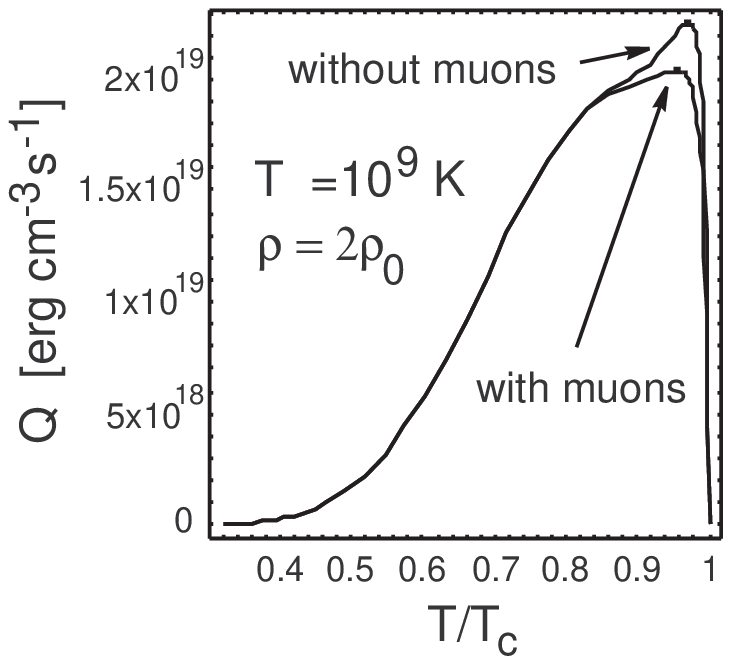} Fig 7. Total neutrino emissivity calculated 
with and without the muon contribution.  
\vskip0.3cm

\section{Efficiency of neutrino emission due to proton pairing}

\label{sec:effic}

The effect of neutron superfluidity and/or proton superconductivity on
different neutrino reactions in the neutron star core is very complicated.
Different neutrino production mechanisms can dominate at different cooling
stages depending on the temperature, and the matter density, and vary along
with the chosen parameters $T_{cp}$, $T_{cn}$, which are the critical
temperatures for the proton and neutron pairing. Fig. 8 shows efficiency of
different reactions at the baryon density $\rho =2\rho _{0}$, where the
Direct Urca process is forbidden. Neutrino emissivities due to the
singlet-state proton pairing, and the triplet-state neutron pairing are
plotted in logarithmic scale against the temperature together with the total
emissivity of two branches of the modified Urca processes, and the total
bremsstrahlung emissivity caused by nn-, np-, and pp-scattering, which are
suppressed due to the neutron superfluidity and/or proton superconductivity
(See \cite{LY93}, \cite{LY94a}, \cite{LY94b}). Two panels of Fig. 8 differ
only by different choice of parameters $T_{cp}$ and $T_{cn}$. On the left
panel we took $T_{cp}=5.6\times 10^{8}$ K, as calculated in \cite{AO85}\ for 
$\beta $-equilibrium nuclear matter, and $T_{cn}=5.6\times 10^{9}$ K, as
obtained in. \cite{HGRR} for triplet-state neutron pairing. On the right
panel we assume $T_{cp}=3.5\times 10^{9}$ K - the result obtained in \cite
{AO85a}, and $T_{cn}=8.5\times 10^{8}$ K, as suggested in \cite{B92}.
\vskip0.3cm 
\psfig{file=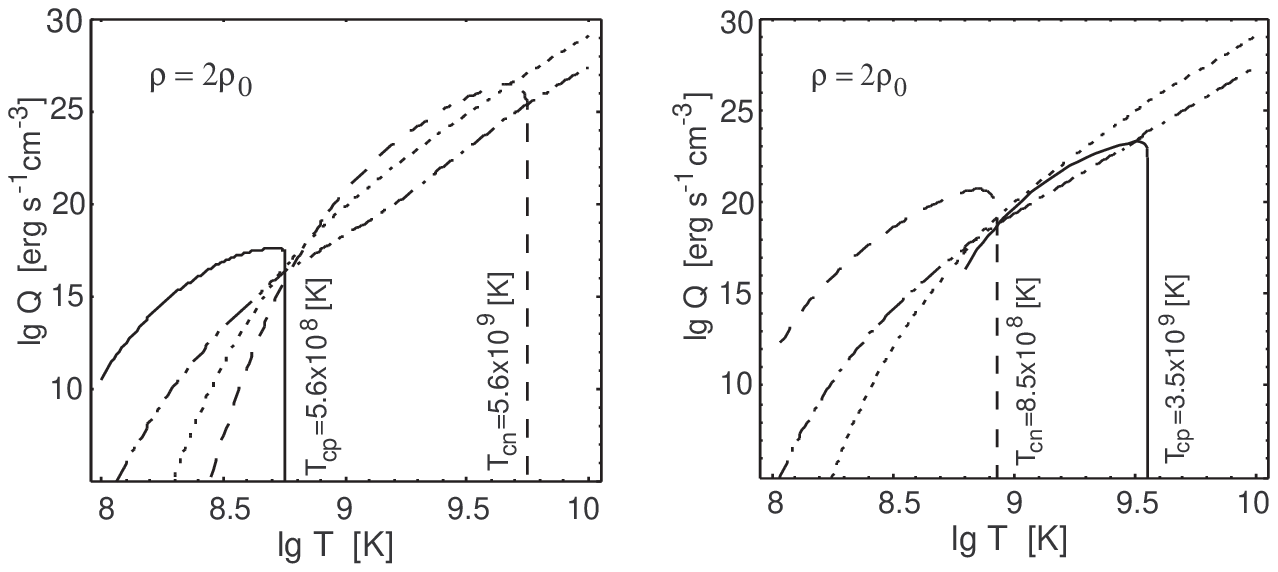} 
Fig 8. Temperature dependence of the neutrino
emissivity in different reactions for $\beta $-equilibrium nuclear matter of
the density $\rho =2\rho _{0}$. The neutrino emissivity due to the
triplet-state neutron pairing is shown by dashed line. The solid line is
the neutrino emissivity due to the singlet-state proton pairing.
Dot-and-dash line shows the total bremsstrahlung emissivity caused by nn-,
np-, and pp-scattering; the dotted line exhibits the total emissivity of two
branches of the modified Urca processes. On the left panel we took $%
T_{cp}=5.6\times 10^{8}$ K, and $T_{cn}=5.6\times 10^{9}$ K. On the right
panel we assume $T_{cp}=3.5\times 10^{9}$ K, and $T_{cn}=8.5\times 10^{8}$ K.
\vskip0.3cm 
\noindent Partial contributions of the proton and neutron
pairing to the total energy losses are very sensitive to the corresponding
critical temperatures. Unfortunately the critical temperatures $T_{cp}$ and $%
T_{cn}$ are not well known up to now. Different authors \cite{AO85} - 
\cite{Ttam93} suggest $T_{cp} $ and $T_{cn}$ which vary in the wide range from 
$10^{8}$ to $10^{10}$
K and sensitively depend on the model of strong interactions used for
calculation. A large scatter of $T_{cp}$ and $T_{cn}$ does not allow to
make choice among many different microscopic results for these parameters.
Therefore, considering the critical temperatures as free parameters, in Fig.
9, we show the regions of $T_{cp}$ and $T_{cn}$, were different neutrino
mechanisms dominate.
\vskip0.3cm 
\psfig{file=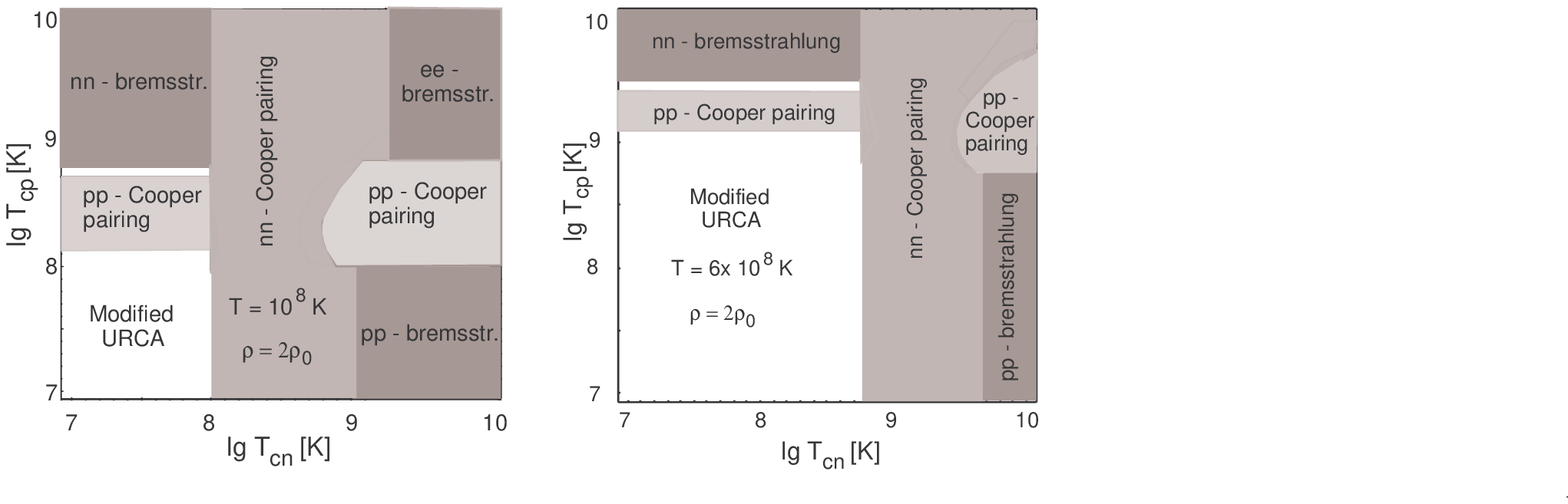}
Fig 9. Regions of $T_{cp}$ and $T_{cn}$,
in which different neutrino reactions dominate at $T=10^{8}$,
 and $6\times 10^{8}$
K in matter of the density  $\rho =2\rho
_{0}$
\vskip0.3cm
\noindent

Additionally to neutrino mechanisms considered above we have included
neutrino bremsstrahlung due to electron-electron collisions recently
calculated by Kaminker and Haensel \cite{KH99}. Unfortunately, their
calculation for neutrino emissivity does not take into account the
collective effects, which substantially modify the effective vector weak
current of the in-medium electron \cite{L99}. However, one can hope that the
reappraised result will be of the same order of magnitude, and the
difference will not be significant in the logarithmic scale we use.

Two panels of Fig. 9 illustrate the case of the standard cooling at $\rho
=2\rho _{0}$ for two internal temperatures of the neutron star: $10^{8}$
and $6\times 10^{8}$ K. Our calculations show that the represented diagrams
does not vary noticeably with the matter density in the region, where the
proton superconductivity exists. Thus, the diagrams of Fig. 9 show the
efficiency of the neutrino production due to proton singlet-state pairing in
the cooling neutron star. One can see that there are wide regions in the
diagrams, where the singlet-state proton pairing dominates in neutrino
production.

\section{Summary and conclusion}

\label{sec:concl}

We have calculated the neutrino-pair emissivity caused by a singlet-state
Cooper pairing of protons by taking into account electromagnetic
correlations among the charged particles in a QED plasma. To incorporate the
electromagnetic correlations we applied the Random phase approximation and
derived Eq. (\ref{QV2}) for  contribution of the vector weak current to
the neutrino emissivity. We found that, virtual electron excitations,
electromagnetically induced by protons undergoing the quantum transition,
generate neutrinos coherently with the initial protons and lead to stronger
neutrino emission than that calculated by different authors without taking
into account the plasma effects. Partial contribution of the proton reaction
to the energy losses from the neutron star is very sensitive to the
critical temperatures for the proton and neutron pairing. Unfortunately the
critical temperatures $T_{cp}$ and $T_{cn}$ are not well known up to now.
Their theoretical values sensitively depend on the model of strong
interactions used for calculation and vary in the wide range from $10^{8}$
to $10^{10}$ K. In Fig. 9, for the temperatures interesting for practice, we
show the regions of $T_{cp}$ and $T_{cn}$, where different neutrino
mechanisms dominate. The represented diagrams does not vary noticeably with
the matter density in the region, where the proton superconductivity exists,
and demonstrate wide domains in the plane $T_{cn}$, $T_{cp}$, where the
singlet-state proton pairing dominates in the neutrino production. 
\acknowledgments 
This work was supported by Spanish Grant DGES PB97- 1432, and the Russian
Foundation for Fundamental Research Grant 00-02-16271.
\newpage
\def\theequation{\Alph{section}.\arabic{equation}}
\setcounter{equation}{0}
\begin{appendix}
\section{Nuclear renormalization of the proton weak vertex}\label{sec:A}

Effect of nucleon-nucleon correlations was evaluated in \cite{52}. Following
this work (Eq. 2.20), the vector weak form-factor of the proton in nuclear
matter is of the following form: 
\begin{equation}
\kappa _{pp}=\tilde{C}_{V}-\frac{f_{np}A_{nn}}{1-2f_{nn}A_{nn}},  \label{29}
\end{equation}
As we use the reduced coupling constants of the proton, the form-factor $%
\kappa _{pp}$ differs by the factor of $1/2$ from that obtained in \cite{52}%
. We have also omited the factors $\tilde{C}_{0}$ and $\tilde{C}_{0}^{-1}$
which finally cancel in the original expression given by the authors. Then,
in the actual case of $k\leq \omega \sim T$, the quantity $A_{nn}$\ can be
written as: 
\begin{equation}
A_{nn}\left( \omega ,k\right) =\frac{1}{2}\left( 1-\frac{s}{2}\ln \frac{s+1}{%
s-1}\right)  \label{30}
\end{equation}
with $s=\omega /kV_{Fn}$. $f_{np}$, $f_{nn}\sim 1$ are constants of the of
the theory of finit Fermi systems.

As the above calculation was made for the case of non-relativistic neutrons
we have to assume that the Fermi velocity of neutrons $V_{Fn}\ll 1.$ Thus,
we have $s\gg 1$. In this limit 
\begin{equation}
A_{nn}\left( \omega ,k\right) \simeq -\frac{1}{6s^{2}}\ll 1  \label{31}
\end{equation}
and the vector weak form-factor of the proton can be simplified as follows 
\begin{equation}
\kappa _{pp}\simeq \tilde{C}_{V}+\frac{f_{np}}{6s^{2}}.  \label{32}
\end{equation}
The constants of the theory of finit Fermi systems are not very good known
for the $\beta $-equilibrium nuclear matter. However, even assuming in the
order of magnitude $f_{np}\sim 1$ we obtain $\kappa _{pp}\sim \tilde{C}_{V}$%
. Thus, the renormalized proton vector weak coupling is found to be much 
smaller than that for electrons. Therefore the nuclear renormalization effects
are negligible with respect to the plasma effects under consideration.
A small nuclear correction to the axial-vector weak coupling of protons can 
also be neglected.
\end{appendix}


\begin{references}
\bibitem{FRS76}  E. Flowers, M. Ruderman, P. Sutherland, ApJ 205 (1976) 541.

\bibitem{Vosk97}  C. Shaab, D. Voskresensky, A. D. Sedrakian, F. Weber, M.
K. Weigel A\&A, 321 (1997){\bf \ }591.

\bibitem{Page98}  D. Page, In: Many Faces of Neutron Stars (eds. R.
Buccheri, J. van Peredijs, M. A. Alpar. Kluver, Dordrecht, 1998) p. 538.

\bibitem{Yak98}  D. G. Yakovlev, A. D. Kaminker, K. P. Levenfish, In:
Neutron Stars and Pulsars (ed. N. Shibazaki et al., Universal Akademy Press,
Tokio, 1998) p. 195.

\bibitem{52}  D. N. Voskresensky, and A. V. Senatorov, Yad. Fiz. 45 (1987)
657 [Sov. J. Nucl. Phys. 45 (1987) 411].

\bibitem{YKL98}  D. G. Yakovlev, A. D. Kaminker, K. P. Levenfish, A\&A 343
(1999) 650.

\bibitem{KHY99}  A. D. Kaminker, P. Haensel, and D. G. Yakovlev, A\&A 345
(1999) L14.

\bibitem{TT93}  T. Takatsuka, R. Tamagaki, Progr. Theor. Phys. Suppl.\ 112
(1993) 27.

\bibitem{L2000}  L. B. Leinson, Phys. Lett. B 473 (2000) 318.

\bibitem{BS93}  E. Braaten and D. Segel, Phys.Rev. D 48 (1993) 1478.

\bibitem{LPkin}  E. M. Lifshitz, L. P. Pitaevskii, Physical Kinetics,
(Pergamon Press, Oxford, 1981).

\bibitem{lp80}  E. M. Lifshitz, L. P. Pitaevskii, Statistical Physics., Part
2 (Pergamon Press, Oxford, 1980).

\bibitem{Mig}  A. B. Migdal, Theory of Finit Fermi Systems and Properties of
Atomic Nuclei [in Russian] ( Nauka, Moscow, 1983); English translation of
earlier eddition: Theory of Finit Fermi Systems (Interscience, New York,
1967).

\bibitem{LY93}  K. P. Levenfish, D. G. Yakovlev, Strongly Coupled Plasma
Physics (Eds H. M. Van Horn, S. Ichimaru, Univ. of Rochester Press,
Rochester, 1993,) p. 167.

\bibitem{LY94a}  K. P. Levenfish, D. G. Yakovlev, Astron. Reports 38 (1994)
247.

\bibitem{LY94b}  K. P. Levenfish, D. G. Yakovlev, Astron. Lett. 20 (1994) 43.

\bibitem{AO85}  L. Amundsen, E. Ostgaard, Nucl. Phys. A 442 (1985) 163.

\bibitem{HGRR}  M. Hoffberg, A. E. Glassgold, R. W. Richardson, M. Ruderman,
Phys. Rev. Lett. 24 (1970) 775.

\bibitem{AO85a}  L. Amundsen, E. Ostgaard, Nucl. Phys. A 437 (1985) 487.

\bibitem{B92}  M. Baldo, J. Cugnon, A. Lejeune, U. Lombardo, Nucl. Phys. A
536 (1992) 349.

\bibitem{Tam70}  R. Tamagaki, Prog. Theor. Phys. 44 (1970) 905.

\bibitem{Ttam93}  T. Takatsuka, R. Tamagaki, Prog. Theor. Phys. 112 (1993)
27.

\bibitem{KH99}  A. D. Kaminker, P. Haensel, Acta Phys. Polonica 30\ (1999)
1125.

\bibitem{L99}  L. B. Leinson, Phys. Lett. B 469{\bf \ }(1999) 166.

\end{references}
\end{document}